\documentclass[%
 reprint,
 amsmath,amssymb,
 prx,
 floatfix
]{revtex4-2}

\usepackage{graphicx}
\usepackage{dcolumn}
\usepackage{bm}
\usepackage{physics}
\usepackage{qcircuit}
\usepackage{amsthm}
\usepackage{mathtools}
\usepackage[colorlinks,linkcolor=blue,anchorcolor=red,citecolor=red]{hyperref}
\usepackage[capitalise]{cleveref}
\usepackage{subcaption}

\let\originalleft\left
\let\originalright\right
\renewcommand{\left}{\mathopen{}\mathclose\bgroup\originalleft}
\renewcommand{\right}{\aftergroup\egroup\originalright}

\DeclarePairedDelimiter\ceil{\lceil}{\rceil}
\DeclarePairedDelimiter\floor{\lfloor}{\rfloor}
\DeclareMathOperator{\spn}{span}
\DeclarePairedDelimiter{\diagfences}{(}{)}
\newcommand{\diag}{\operatorname{diag}\diagfences}
\DeclarePairedDelimiter{\nint}\lfloor\rceil

\begin{document}

\title{Deterministic spatial search using alternating quantum walks}

\author{S. Marsh}
\email{samuel.marsh@research.uwa.edu.au}
\affiliation{Department of Physics, The University of Western Australia, Perth, Australia}

\author{J. B. Wang}
\email{jingbo.wang@uwa.edu.au}
\affiliation{Department of Physics, The University of Western Australia, Perth, Australia}

\date{April 12, 2021}


\begin{abstract}
This paper examines the performance of spatial search where the Grover diffusion operator is replaced by continuous-time quantum walks on a class of interdependent networks. We prove that for a set of optimal quantum walk times and marked vertex phase shifts, a deterministic algorithm for structured spatial search is established that finds the marked vertex with 100\% probability. This improves on the Childs \& Goldstone spatial search algorithm on the same class of graphs, which we show can only amplify to 50\% probability. Our method uses $\ceil{\frac{\pi}{2\sqrt{2}}\sqrt{N}}$ marked vertex phase shifts for an $N$-vertex graph, making it comparable with Grover's algorithm for unstructured search. It is expected that this new framework can be readily extended to deterministic spatial search on other families of graph structures.
\end{abstract}

\maketitle

\section{Introduction}

Quantum search on a spatially structured database via a continuous-time quantum walk (CTQW) is an important and broad problem in quantum computation \cite{Mlken2011,Childs2004,Magniez2011,Aaronson,VenegasAndraca2012}, which aims to find an unknown marked vertex on an underlying graph of specified topology. Many spatial structures studied so far admit a quadratic quantum speedup for continuous-time spatial search using the Childs \& Goldstone ($\mathcal{CG}$) algorithm \cite{Childs2004,Janmark2014,Chakraborty2016,Wong_2016,Cattaneo2018,Chakraborty2020}, where the probability of measuring the marked vertex $\ket{\omega}$ can be amplified to $\mathcal{O}(1)$ in $\mathcal{O}(\sqrt{N})$ time for an $N$-vertex graph. However, with the exception of the complete graph, the vast majority of graphs studied previously have a success probability less than 1 \cite{Chakraborty2020}.

More recently, with the significant research interests in the Quantum Approximate Optimisation Algorithm (QAOA) \cite{farhi2014quantum,a12020034}, searching for a marked element has been treated variationally as a combinatorial optimisation problem having the objective function
\begin{equation}
    f(x) = \begin{cases}
    1 & \text{$x$ is the marked element $\omega$,} \\
    0 & \text{otherwise.}
    \end{cases}
\end{equation}
In QAOA-based algorithms the aim is to optimise the parameters of a quantum circuit to maximise the expectation value of the output state with respect to the objective function. Specifically, for database search, \citet{Morales2018} show that Grover's algorithm can be learnt variationally with a parameterised Grover diffusion operator, and \citet{Jiang2017} demonstrate that replacing the diffusion operator with a parameterised transverse field operator can also result in an efficient search algorithm. 

These results for two different so-called `mixing operators' suggest a generalisation for a gate-model search framework. Motivated by the fact that the above mixers are equivalent to a CTQW on the complete graph \cite{Morales2018} and hypercube \cite{Jiang2017} respectively, we propose a new spatial search framework based on the state evolution
\begin{equation}
    \ket{\Vec{t}, \Vec{\theta}} = \left(\prod\limits_{j=1}^p U_w(t_j) U_f(\theta_j) \right)\ket{s} \, .
\end{equation}
Here, starting in the equal superposition $\ket{s}$, applications of $U_f(\theta) = e^{-i \theta \ket{\omega} \bra{\omega}}$ perform a controlled phase shift $\theta$ to the marked vertex, and each $U_w(t)=e^{- i t  A}$ applies a quantum walk for time $t$ over a graph having adjacency matrix $A$. Our aim is to determine values for $\Vec{t}$ and $\Vec{\theta}$ that maximise the overlap with the marked element after $p$ iterations. This is a specific case of the Quantum Walk Optimisation Algorithm applied to search \cite{Marsh2020,Marsh2019}.



Expressing the spatial search problem in this framework has a number of benefits. As in \cite{Jiang2017}, it can result in a mixing circuit that uses substantially fewer gates than the standard Grover diffusion operator, having advantages for NISQ hardware. More generally, hardware with limited qubit connectivity can benefit from a mixing operator that suits the couplings. In addition, in terms of quantum optimisation, studying the query complexity of a particular mixer on the search problem gives an analytical baseline for the mixer's performance on more complex combinatorial optimisation problems. Finally, there are interesting advantages over the $\mathcal{CG}$ spatial search scheme \cite{Childs2004}. As well as being naturally suited to gate model circuit implementation, for the class of graphs we study in this paper deterministic search can be achieved using the alternating phase-walk formulation. This is, to the authors' knowledge, the first example of deterministic search on a spatially structured database that goes beyond the direct use of the generalised Grover diffusion operator \cite{Long2001,Brassard2002}. It improves on the $\mathcal{CG}$ algorithm on the same class of graphs, which we will show only reaches 50\% success probability.

\begin{figure}
    \centering
    \includegraphics[width=0.9\linewidth]{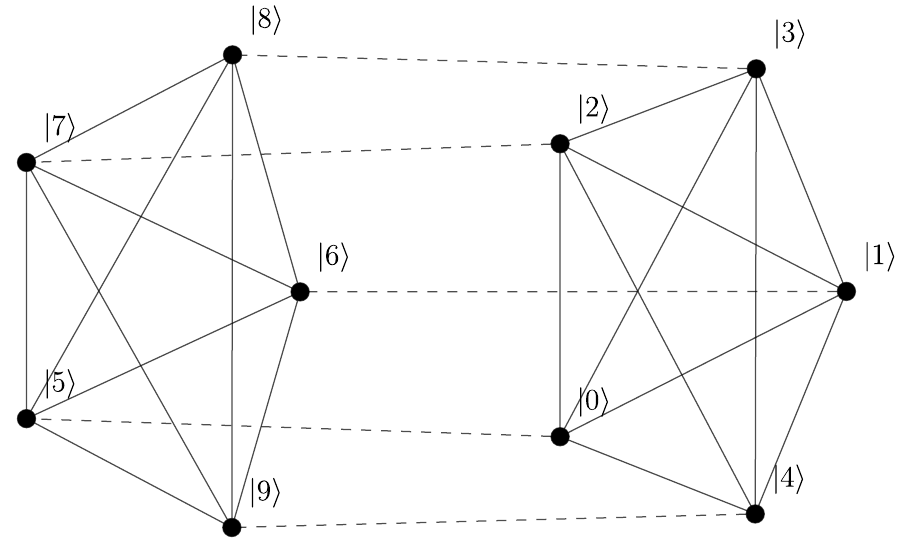}
    \caption{The $n=5$ CIIN has a total of $N=2n=10$ vertices. The solid lines are the edges of the complete graph $\mathbb{K}_5$, whilst the dashed lines are the identity interconnections.}
    \label{fig:cylinder}
\end{figure}

In this paper, we focus on finding closed-form expressions for parameters $(\Vec{t}, \Vec{\theta})$ for a particular class of interdependent networks to achieve efficient deterministic quantum spatial search. A $2n$-vertex interdependent network $G$ is composed via a block adjacency matrix
\begin{equation}
    A_\text{full} = \begin{pmatrix}
        A_1 & A_2 \\
        A_2^T & A_1
    \end{pmatrix} \, ,
\end{equation}
where $A_1$ and $A_2$ are the $n \times n$ adjacency matrices of the (undirected) primary and interlink graphs $G_1$ and $G_2$ respectively \cite{VanMieghem2016}. Here, $A_2$ provides the connectivity between two copies of $A_1$. Specifically, we focus on the class of interdependent networks where $G_1 = \mathbb{K}_n$ is the complete graph, and $G_2 = \mathbb{I}_n$ is an identity interconnection graph. We refer to this class of graphs as CIINs (Complete Identity Interdependent Networks). A $10$-vertex example is shown in \cref{fig:cylinder}, where it is clear that the graph takes a `cylindrical' form with each vertex $\ket{j}$ having an opposite vertex $\ket{(j + n) \mod 2 n}$. The CIIN is equivalent to a $n \times 2$ Rook graph, where the general case of a $n \times m$ Rook graph is studied by \citet{Chakraborty2020} in the context of the $\mathcal{CG}$ framework.

This class of graphs is particularly interesting for study in the context of comparing standard quantum spatial search to the proposed alternating phase-walk framework, due to the disparity between standard spatial search and the proposed framework. In \cref{sec:walksubspace}, we introduce the CIIN `walk subspace' to which the search dynamics can be reduced, for the purposes of simplified analysis. In \cref{sec:spatialsearch}, we examine the performance of the $\mathcal{CG}$ algorithm on CIINs as a basis for comparison, which achieves 50\% success probability on CIINs after time $T=\frac{\pi}{2\sqrt{2}}\sqrt{N}$. In \cref{sec:alternatingalg}, we present the `alternating phase-walk' algorithm and analyse its properties.
We then discuss gate-model implementation of the algorithm in \cref{sec:circuit}. We conclude with general discussion of the algorithm in \cref{sec:discussion}.

\section{Walk subspace}
\label{sec:walksubspace}

Although quantum spatial search takes place in the full $N$-dimensional Hilbert space, due to high symmetry in the CIIN graph the dynamics can be reduced analytically to the consideration of a 4-dimensional `walk subspace'. In addition, since CIINs are vertex-transitive, for the purposes of analysis we can set the marked vertex as $\ket{\omega} = \ket{0}$ and the resulting algorithm will apply regardless to an arbitrary marked vertex. Consequently, relative to the marked vertex $\ket{0}$, there are four distinct categories of vertex that together define the walk basis $B=\{\ket{b_1}, \ket{b_2}, \ket{b_3}, \ket{b_4}\}$:
\begin{itemize}
    \item the marked node $\ket{b_1} = \ket{\omega} = \ket{0}$,
    \item the opposite node $\ket{b_2} = \ket{\tilde{\omega}} = \ket{n}$,
    \item the equal superposition over the unmarked vertices on the same side as $\ket{\omega}$, \begin{equation}
        \ket{b_3} = \frac{1}{\sqrt{n-1}} \sum\limits_{x=1}^{n-1} \ket{x} \, ,
    \end{equation}
    \item and the equal superposition over the vertices on the same side as $\ket{\tilde{\omega}}$, excluding $\ket{\tilde{\omega}}$,
    \begin{equation}
        \ket{b_4} = \frac{1}{\sqrt{n-1}} \sum\limits_{x=n+1}^{N-1} \ket{x} \, .
    \end{equation}
\end{itemize}
The above basis can be obtained simply by symmetry arguments, or alternatively by applying systematic dimensionality reduction via the Lanczos algorithm as per \cite{Novo2015}.

We initialise the system in the equal superposition, which in the reduced walk basis has the form
\begin{equation}
        \ket{s} = \frac{1}{\sqrt{2n}}\left(1,1,\sqrt{n-1},\sqrt{n-1}\right) \, ,
\end{equation}
where approximately half the probability is contained in each of the last two vertex groups. A proof for the adjacency matrix taking this form can be found in \cref{sec:reduced-adj}.

\begin{figure}
    \centering
    \includegraphics[width=0.8\linewidth]{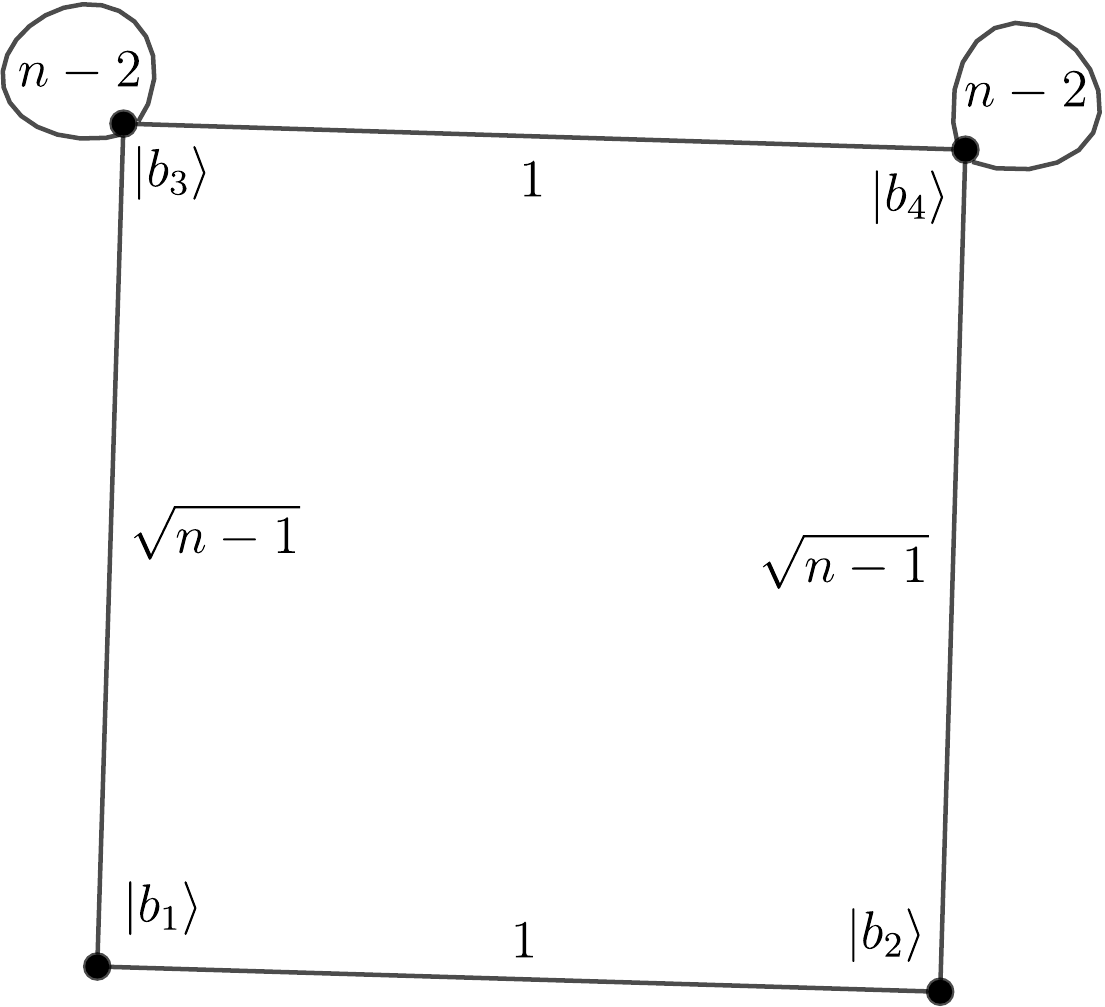}
    \caption{The weighted graph representing connectivity between the four vertex groups with respect to the marked vertex $\ket{b_1} = \ket{\omega}$.}
    \label{fig:reduced-graph}
\end{figure}

To determine the reduced adjacency matrix $A_\text{full} \mapsto A$, $\bra{b_i} A_\text{full} \ket{b_j}$ is calculated to obtain
\begin{align}
    A = \begin{pmatrix}
 0 & 1 & \sqrt{n-1} & 0 \\
 1 & 0 & 0 & \sqrt{n-1} \\
 \sqrt{n-1} & 0 & n-2 & 1 \\
 0 & \sqrt{n-1} & 1 & n-2 \\
\end{pmatrix} \, .
\label{eq:adjwalkbasis}
\end{align}
This reduced adjacency matrix can itself be treated as an undirected but weighted graph as illustrated in \cref{fig:reduced-graph}.

\section{Applying $\mathcal{CG}$ spatial search}
\label{sec:spatialsearch}

We first briefly explore the performance of the $\mathcal{CG}$ algorithm \cite{Childs2004} on CIINs, in order to form a basis for comparison with our algorithm. The results in this section are consistent with the general analysis of optimality conditions for the $\mathcal{CG}$ algorithm given in \cite{Chakraborty2020}.
In $\mathcal{CG}$ spatial search, the system is evolved from the equal superposition under the Hamiltonian
\begin{equation}
    H = -\gamma A - \ket{\omega} \bra{\omega}
\end{equation}
where $A$ is the graph adjacency matrix and $\ket{\omega}\bra{\omega}$ is the oracular marking Hamiltonian. The Laplacian is often used in place of the adjacency matrix, however, CIINs are regular graphs so the dynamics are equivalent up to a global phase. The aim for $\mathcal{CG}$ spatial search is to find a critical value of the variational parameter $\gamma=\gamma^*$ that will induce `fast' rotation between the initial superposition and a state having $\mathcal{O}(1)$ overlap with the marked state.

In the $\mathcal{CG}$ spatial search framework, we define our Hamiltonian as
\begin{align}
    H &= -\gamma A - \ket{\omega} \bra{\omega} \\
    &= -\gamma
\begin{pmatrix}
 \frac{1}{\gamma } & 1 & \sqrt{n-1} & 0 \\
 1 & 0 & 0 & \sqrt{n-1} \\
 \sqrt{n-1} & 0 & n-2 & 1 \\
 0 & \sqrt{n-1} & 1 & n-2 \\
\end{pmatrix} \\
&\approx -\gamma 
\begin{pmatrix}
 \frac{1}{\gamma } & 1 & \sqrt{n} & 0 \\
 1 & 0 & 0 & \sqrt{n} \\
 \sqrt{n} & 0 & n & 1 \\
 0 & \sqrt{n} & 1 & n \\
\end{pmatrix}
\end{align}
for large $n$.

We follow convention and apply degenerate perturbation theory to find the critical value $\gamma^*$. Assuming $\gamma = \mathcal{O}(\frac{1}{N})$ and separating the Hamiltonian by terms having the same order of magnitude,
\begin{equation}
    H = H_0 + H_1 + \mathcal{O}\left(\frac{1}{N}\right) \, ,
\end{equation}
where
\begin{align}
H_0 &= \begin{pmatrix}
 -1 & 0 & 0 & 0 \\
 0 & 0 & 0 & 0 \\
 0 & 0 & -\gamma n & 0 \\
 0 & 0 & 0 & -\gamma n \\
\end{pmatrix} , \\
H_1 &= 
\begin{pmatrix}
 0 & 0 &  -\gamma\sqrt{n} & 0 \\
 0 & 0 & 0 &   -\gamma\sqrt{n} \\
 -\gamma \sqrt{n} & 0 & 0 & 0 \\
 0 & -\gamma \sqrt{n} & 0 & 0 , \\
\end{pmatrix} \, .
\end{align}
Here $H_0$ consists of terms of order $\mathcal{O}(1)$, $H_1$ of order $\mathcal{O}(\frac{1}{\sqrt{N}})$, and we drop higher-order terms.

Recall that the initial state is approximately $\ket{s} \approx \frac{1}{\sqrt{2}}\left( \ket{b_3} + \ket{b_4}\right)$ for large $n$. Thus in order to induce an approximate rotation between the initial superposition and the marked state, we set $-1 = -\gamma^*n \implies \gamma^* = \frac{1}{n}$. 

\begin{figure}
    \centering
    \includegraphics[width=0.95\linewidth]{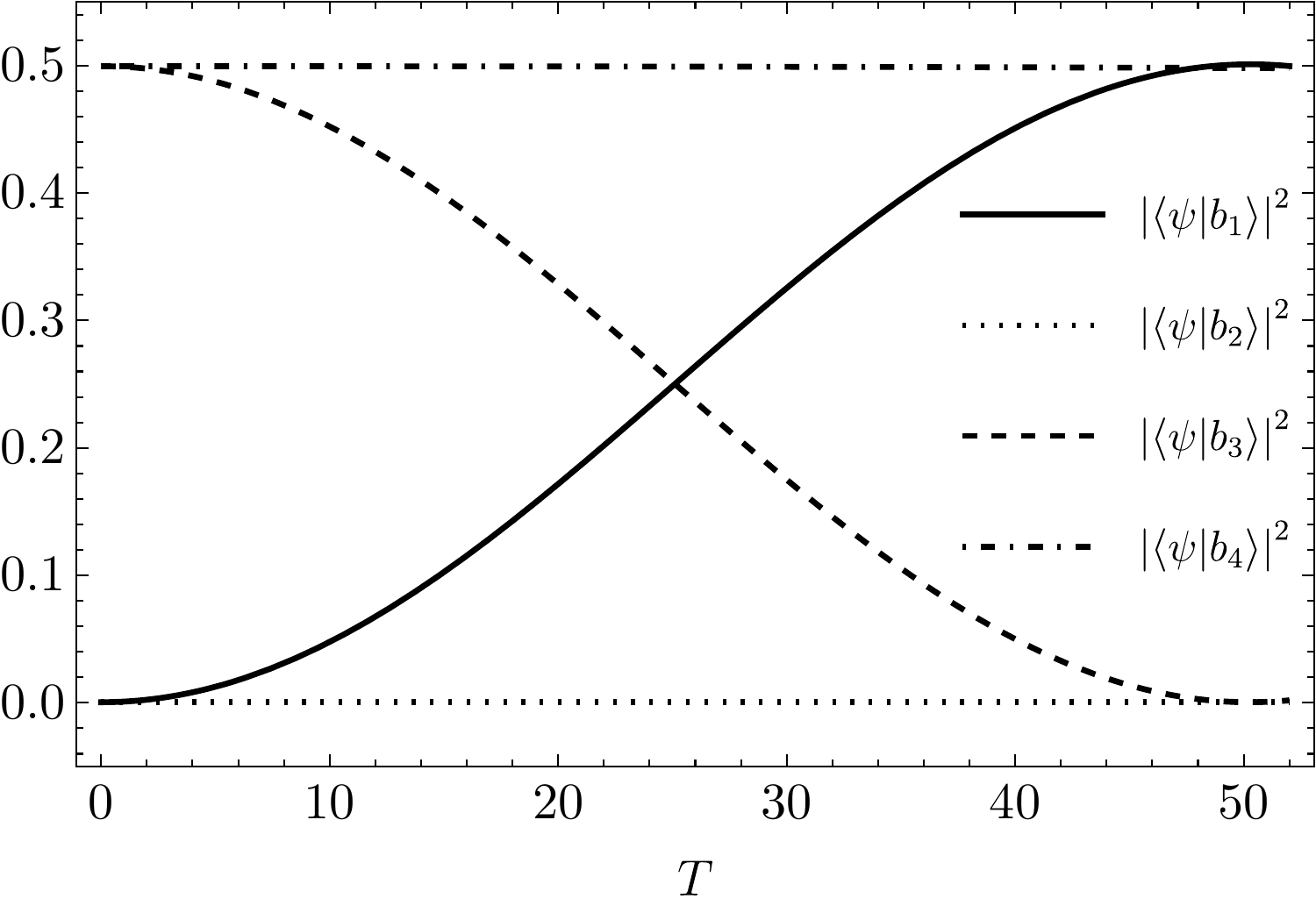}
    \caption{Dynamics of the spatial search algorithm on a CIIN with $N=2048$ vertices, showing the rotation between $\ket{b_3}$ and the marked element $\ket{b_1} = \ket{\omega}$. Approximately 50\% success probability is reached after evolution time $T=\frac{\pi}{2\sqrt{2}}\sqrt{N} \approx 50.3$.}
    \label{fig:spatialsearch-cylinder-1024}
\end{figure}

Solving the resulting system $H_0 + H_1$ with this critical value, the relevant eigenvalues are $-1 \pm \frac{1}{\sqrt{n}}$ with corresponding eigenstates $\frac{1}{\sqrt{2}} \left( \ket{b_3} \mp \ket{b_1} \right)$
and thus a rotation is induced (approximately) between the marked state and the third basis element. The eigenvalue difference is $\Delta E = \frac{2}{\sqrt{n}}$ and thus after time $T=\frac{\pi}{\Delta E} = \frac{\pi}{2}\sqrt{n}=\frac{\pi}{2\sqrt{2}}\sqrt{N}$ we have increased the probability of measuring the marked element to about half. A numerical simulation of the $\mathcal{CG}$ spatial search algorithm on a $2048$-vertex CIIN is shown in \cref{fig:spatialsearch-cylinder-1024}, where it is clear that the rotation is induced between $\ket{b_1} \leftrightarrow \ket{b_3}$ with the other two basis states remaining unchanged.

We can intuitively interpret the action of the $\mathcal{CG}$ spatial search algorithm on CIINs as inducing the amplitude from vertices on the same side as the marked element to `flow' into the marked element, while leaving the other side of the graph unchanged. Although this algorithm satisfies the criterion for efficient quantum spatial search (reaching $\mathcal{O}(1)$ success probability after $\mathcal{O}(\sqrt{N})$ time), it is somewhat unsatisfying that the search algorithm is `ignoring' half of the interdependent network. One could heuristically argue that the well-known result of $\sim 100\%$ probability for $\mathcal{CG}$ spatial search on the complete graph is being `exploited' on half of the CIIN, rather than the CIIN itself fundamentally admitting efficient $\mathcal{CG}$ spatial search.


\section{Alternating walk algorithm}
\label{sec:alternatingalg}

In this section, we present a new approach to evolve the equal superposition to the marked vertex state. We will introduce the core algorithm, which analogous to Grover's algorithm has failure rate going to zero as $n \rightarrow \infty$. Then, in the same vein as the Grover-Long algorithm \cite{Long2001}, we show that the search can be made fully deterministic by parametrising the phase shifts.

\subsection{Quantum walk on CIIN}

\begin{figure}[b]
    \centering
    \includegraphics[width=0.9\linewidth]{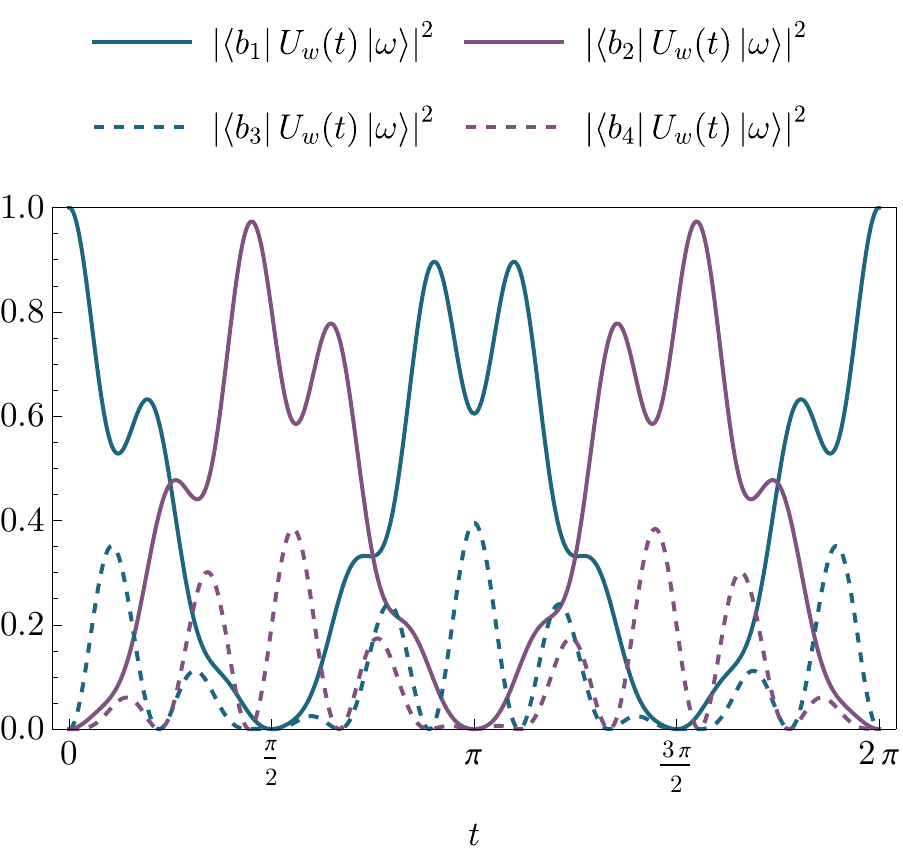}
    \caption{Dynamics of a quantum walk on an 18-vertex CIIN, starting from the marked element. There is $2\pi$-periodicity, and for multiple values of $t$ the amplitude is suppressed in one or more vertex groups.}
    \label{fig:walkfrommarked}
\end{figure}
First, we examine some essential properties of quantum walks on CIINs.
Quantum walk on a CIIN is $2\pi$-periodic, since the graph is regular with integer adjacency matrix eigenvalues -2, 0, $(n-2)$ and $n$ \cite{Godsil2011}. We observe the state of the system after a quantum walk is applied starting in the single marked vertex, $U_w(t) \ket{\omega}$. The overlap of each basis state after a quantum walk for an 18-vertex example is shown in \cref{fig:walkfrommarked}. It is worth to note that for any $k \in \mathbb{Z}$,
\begin{equation}
    U_w(\frac{2 \pi k}{n}) \ket{\omega} = \cos(\frac{2 \pi  k}{n}) \ket{\omega} - i \sin(\frac{2 \pi  k}{n}) \ket{\Tilde{\omega}} \, .
    \label{eq:walkfrommarked}
\end{equation}
That is, for specific walk times a vertex only interacts with its opposite vertex, with probability in the other vertex groups destructively interfering. We also observe from \cref{eq:walkfrommarked} that the probability in the opposite vertex is maximised when $k = \nint{n/4}$. Clearly, the graph exhibits perfect state transfer \cite{Godsil2012} from one vertex to its opposite when $n \bmod 4 = 0$.

\subsection{The dual basis}

The core of the CIIN search algorithm is to rotate between eigenstates of the adjacency matrix, spanning a two-dimensional subspace of the walk space. Hence for the purposes of a detailed mathematical analysis, we introduce a new basis consisting of the eigenstates of $A$. These basis elements are simply the eigenvectors of the original (and significantly, unperturbed) adjacency matrix in the walk basis, which are straightforward to compute:
\begin{align}
    \ket{b_1^*} & = \frac{1}{\sqrt{2n}}\left(1, 1, \sqrt{n-1}, \sqrt{n-1}\right) = \ket{s} \, ,\\
    \ket{b_2^*} &= \frac{1}{\sqrt{2n}}\left( -1, 1, -\sqrt{{n-1}}, \sqrt{{n-1}}\right) \, ,\\
    \ket{b_3^*} &= \frac{1}{\sqrt{2n}}\left( \sqrt{{n-1}}, -\sqrt{{n-1}}, -1, 1 \right) \, ,\\
    \ket{b_4^*} &= \frac{1}{\sqrt{2n}}\left( -\sqrt{{n-1}}, -\sqrt{{n-1}}, 1, 1 \right) \, .
\end{align}
The equal superposition is an eigenvector of any regular graph, and hence is the first basis state in the dual basis. In this basis, the marked state takes the form
\begin{align}
    \ket{\omega} &= \frac{1}{\sqrt{2n}}\left( 1, -1, \sqrt{{n-1}}, -\sqrt{{n-1}}\right) \, ,
\end{align}
and the initial state is
\begin{align}
        \ket{s} &= \left( 1, 0, 0, 0 \right) \, .
\end{align}
We then have
\begin{align}
    U_w(t) &= \exp{-i t \diag{n, n-2, -2, 0}} \, ,\\
    U_f(\theta) &= \exp{-i \theta \ket{\omega} \bra{\omega}} \, .
\end{align}
Thus in the dual basis the two unitaries take opposite roles: in contrast to the original walk basis, here the quantum walk only applies a phase difference to the basis states, whilst the marking unitary now performs the `mixing'.

\subsection{Approximate algorithm}

The algorithm aims to rotate the system from the initial superposition $\ket{s}$ to the entangled state
\begin{equation}
    \ket{+_\omega} = \frac{1}{\sqrt{2}}\left( \ket{\omega} + \ket{\tilde\omega}\right) = \frac{1}{\sqrt{2}}\left( \ket{b_1} + \ket{b_2} \right) \, .
\end{equation}
To design an appropriate iteration, observe that in the dual basis,
\begin{equation}
    \ket{+_\omega} = \frac{1}{\sqrt{n}} \ket{b_1^*} - \sqrt{\frac{n-1}{n}}\ket{b_4^*} \ ,
    \label{eq:entangled-dual}
\end{equation}
so the system should be rotated from the initial state towards the state $\ket{b_4^*}$.
In order to achieve this, consider the parametrised unitary
\begin{equation}
    U(t, k) = U_w(t) U_f(\pi) U_w(\frac{2\pi k}{n}) U_f(\pi)
\end{equation}
where $k$ is an integer. In the dual basis, we find that when $j= 2, 3$,
\begin{align}
    \bra{b_j^*} U(t,k) \ket{b_1^*} = \bra{b_1^*} U(t,k) \ket{b_j^*}= 0 \\
    \bra{b_j^*} U(t,k) \ket{b_4^*} = \bra{b_4^*} U(t,k) \ket{b_j^*}= 0
\end{align}
and thus $U(t, k)$ restricts to the desired 2-dimensional subspace $\{ \ket{b_1^*}, \ket{b_4^*} \}$. Consequently, we define the iterate
\begin{equation}
    U = U_w(t_2) U_f(\pi) U_w(t_1) U_f(\pi)
\end{equation}
where
\begin{align}
    t_1 &= \frac{2 \pi}{n} \nint{\frac{n}{4}} \, , \\
    t_2 &= -\frac{2}{n} \arctan\left( \frac{n-2}{n} \tan t_1 \right) \, .
\end{align}
These parameters to $U(t, k)$ are chosen to maximise $\abs{\bra{b_4^*}U(t, k)^2\ket{b_1^*}}^2$, i.e. to maximise the rotation towards the $\ket{b_4^*}$ state.
For large $n$, the parameters converge to $t_1 \approx \frac{\pi}{2}$ and $t_2 \approx \frac{\pi}{n}$.

In the 2-dimensional subspace $\{\ket{b_1^*}, \ket{b_4^*}\}$, the iterate takes the form
\begin{equation}
    U = \begin{pmatrix}
 \frac{n+e^{2 i t_1}-1}{n} & -\frac{\sqrt{n-1} \left(-1+e^{2 i t_1}\right)}{n} \\
 -\frac{\sqrt{n-1} \left(-1+e^{2 i t_1}\right) e^{i n t_2}}{n} & \frac{\left(1+(n-1) e^{2 i t_1}\right) e^{i n t_2}}{n} \\
\end{pmatrix} \, .
\end{equation}
If a global phase is introduced such that $\arg \bra{b_1^*} U \ket{b_1^*} = 0$, we find that this matrix has eigenphases
\begin{equation}
    \lambda_\pm = \pm \arcsin\left( \frac{2 \sqrt{n-1}}{n} \sin t_1 \right)
\end{equation}
with corresponding eigenstates
\begin{equation}
    \ket{v_\pm} = \frac{1}{\sqrt{2}} \left(\ket{b_4^*} \mp e^{-i n t_2 / 2}\ket{b_1^*}\right) \, .
\end{equation}

\begin{figure}
    \centering
    \includegraphics[width=0.9\linewidth]{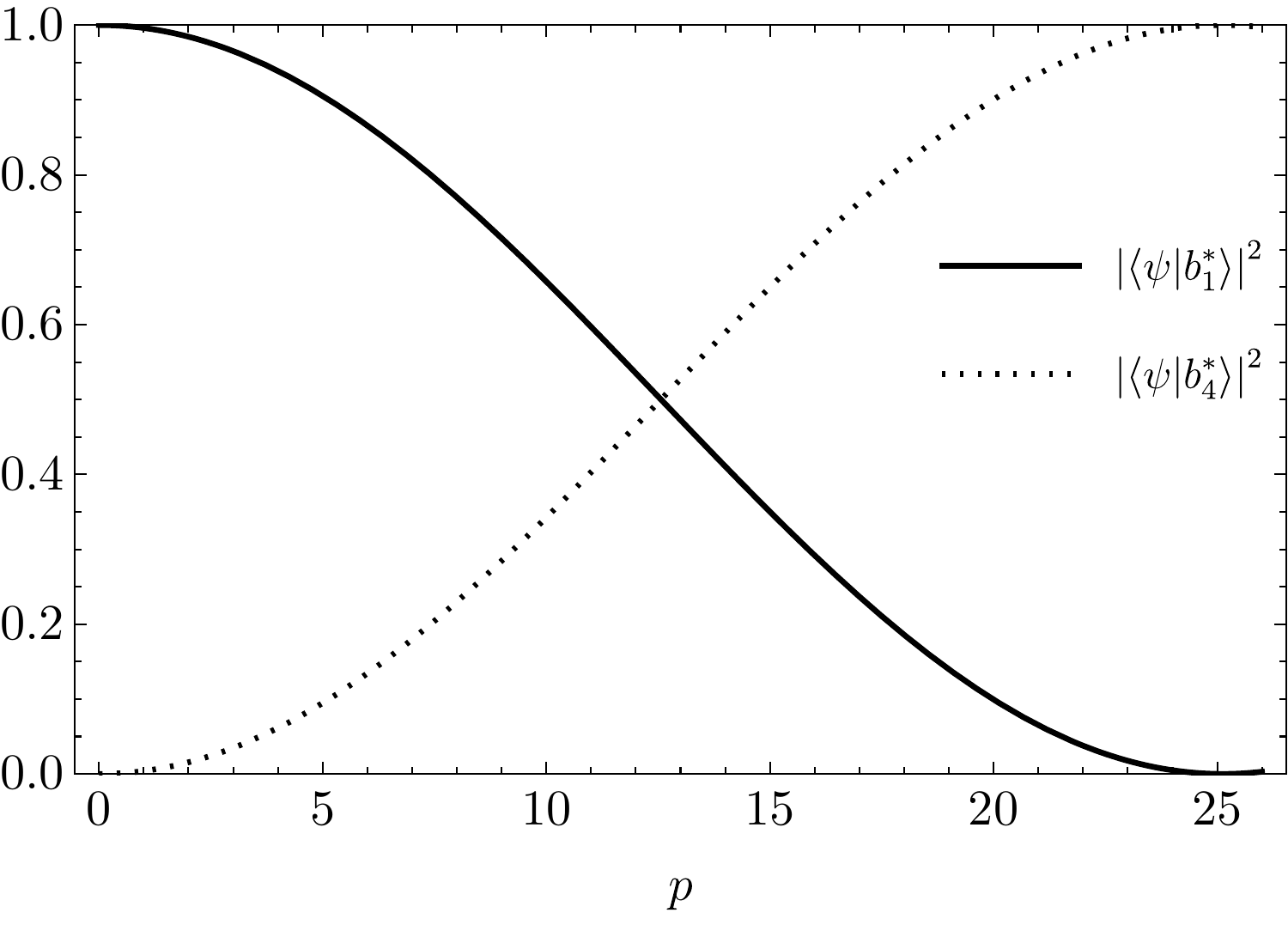}
    \caption{The state of a $2048$-vertex CIIN after $p$ applications of $U$ to $\ket{s}$, shown in the dual basis.}
    \label{fig:evolution-dual-basis-2048}
\end{figure}

Using the eigensystem of $U$ to diagonalise and compute the matrix power,
\begin{equation}
    U^p \ket{b_1^*} = \cos\left( p \lambda_+ \right) \ket{b_1^*} - i e^{i n t_2 / 2} \sin\left( p \lambda_+ \right) \ket{b_4^*} \, .
\end{equation}
Hence, when $p=\frac{1}{\lambda_+} \arccos{\frac{1}{\sqrt{n}}}$ we reach the state
\begin{equation}
    \ket{\psi} = \frac{1}{\sqrt{n}} \ket{b_1^*} - i e^{i n t_2 / 2} \sqrt{\frac{n-1}{n}}\ket{b_4^*} \, ,
\end{equation}
which is nearly the desired entangled state $\ket{+_\omega}$ given in \cref{eq:entangled-dual}, but with a different phase on $\ket{b_4^*}$. Observe that applying a final quantum walk for time $t_3$ to this state gives
\begin{equation}
    U_w(t_3) \ket{\psi} = \frac{1}{\sqrt{n}} \ket{b_1^*} -i e^{\frac{1}{2} i n \left(t_2+2 t_3\right)} \sqrt{\frac{n-1}{n}}\ket{b_4^*} \, ,
\end{equation}
and thus with $t_3 = \frac{\pi }{2 n}-\frac{t_2}{2}$ the state $\ket{\omega_+}$ is reached. We illustrate a numerical simulation of this evolution for a $2048$-vertex system in \cref{fig:evolution-dual-basis-2048}.

Reaching the entangled state $\ket{+_\omega}$ is sufficient to solve the search problem with 100\% success as $N \rightarrow \infty$. There are two approaches:
\begin{enumerate}
    \item One can measure the system to obtain either the marked vertex or the vertex `opposite' it: the marking oracle can be queried one more time to confirm which. If the system is measured to be in state $\ket{x}$, where $x$ is then verified to \textit{not} be the marked vertex, then $\ket{\omega} = \ket{x + n \mod N}$.
    \item For a fully coherent algorithm, observe from \cref{eq:walkfrommarked} that when $k=\nint{\frac{n}{8}}$,
    \begin{equation}
        U_w(\frac{2 \pi k}{n}) \ket{\omega} \approx \frac{1}{\sqrt{2}}\left( \ket{\omega} - i \ket{\tilde{\omega}} \right) \, .
    \end{equation}
    Thus, $\ket{\omega} \approx U_w(\frac{2 \pi}{n}\nint{\frac{n}{8}}) U_f(\frac{\pi}{2}) \ket{+_\omega}$.
\end{enumerate}

Consequently, this algorithm achieves essentially 100\% probability (as $n$ increases) of measuring the marked element after
\begin{equation}
    \frac{2}{\lambda_+}\arccos{\frac{1}{\sqrt{n}}} + 1 \approx \frac{\pi}{2}\sqrt{n} = \frac{\pi}{2\sqrt{2}}\sqrt{N}
\end{equation}
oracle queries, which is a factor of $\sqrt{2}$ away from the optimal $\frac{\pi}{4}\sqrt{N}$ queries used by Grover's algorithm.

The following section will show how the algorithm can be made fully deterministic to achieve 100\% theoretical success probability.

\subsection{Deterministic algorithm}

\begin{figure*}
    \centering
    \begin{subfigure}[t]{0.48\textwidth}
        \centering
        \includegraphics[width=\textwidth]{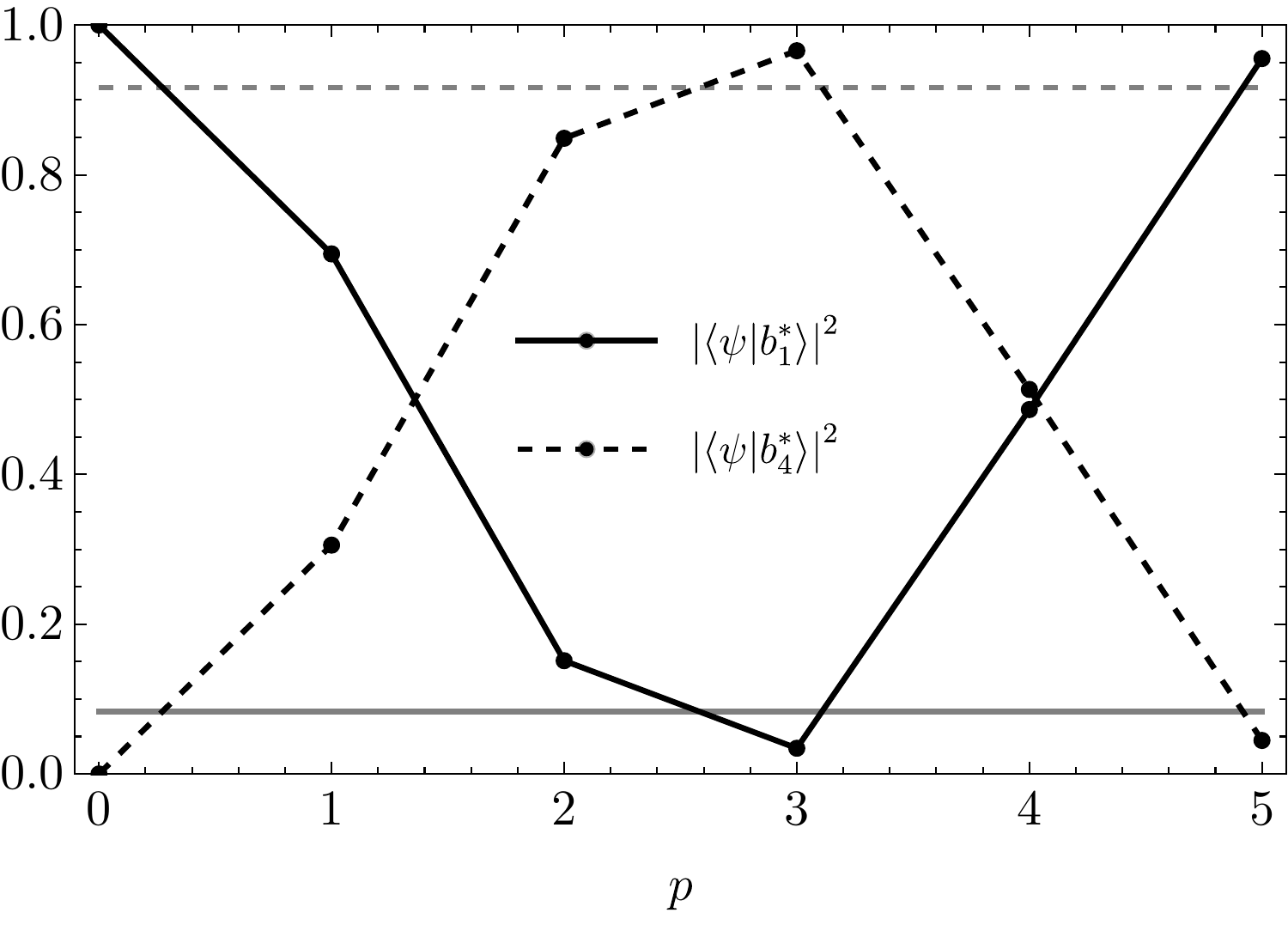}
        \caption{}
    \end{subfigure}
    ~ 
    \begin{subfigure}[t]{0.48\textwidth}
        \centering
        \includegraphics[width=\textwidth]{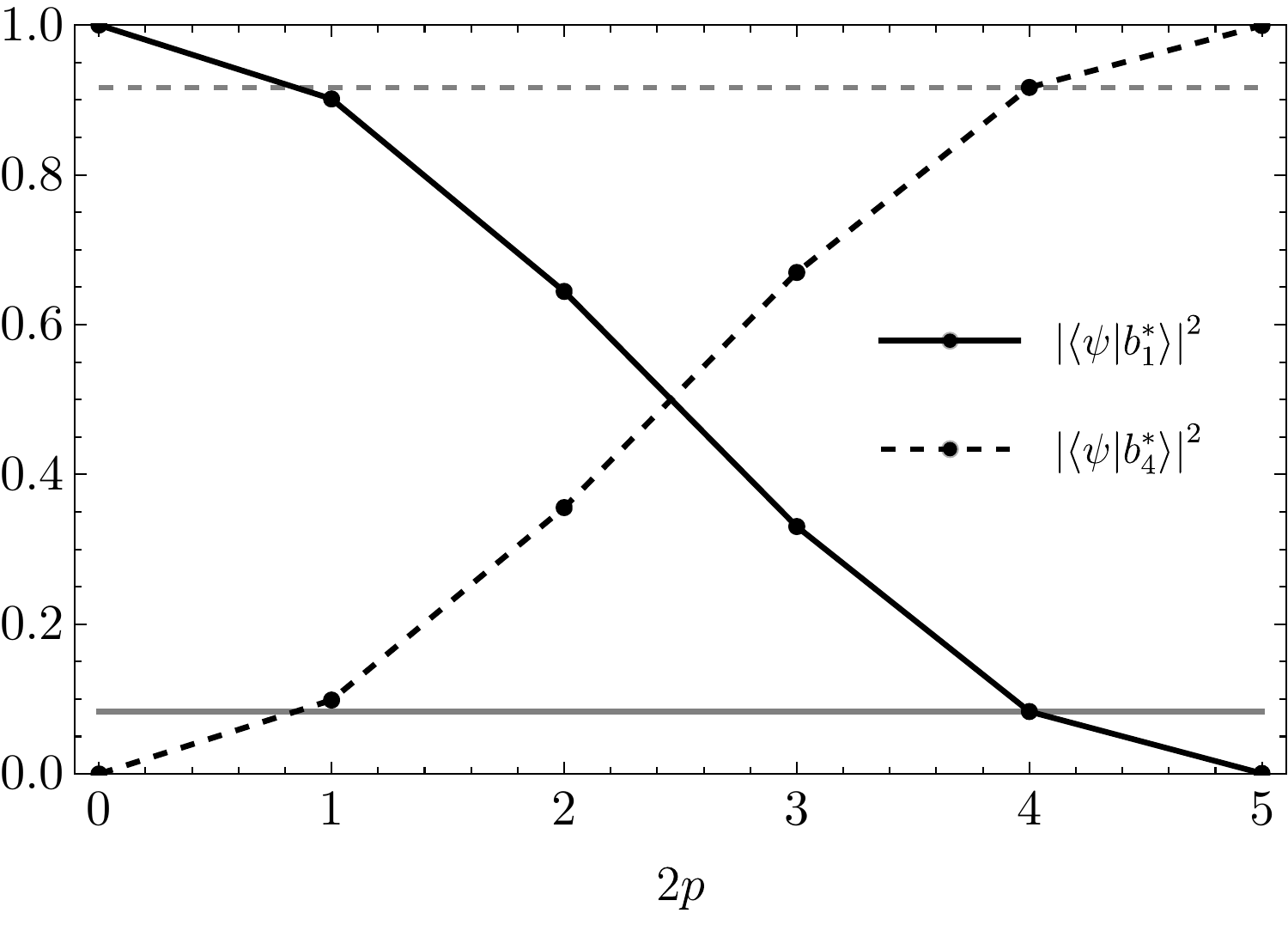}
        \caption{}
    \end{subfigure}
    \caption{Comparing the dynamics of the (a) approximate and (b) deterministic algorithms on a 24-vertex instance. The horizontal lines represent the evolution target of $\abs{\braket{\psi}{b_1^*}}^2 = \frac{1}{n}$ (and $\abs{\braket{\psi}{b_4^*}}^2 = \frac{n-1}{n}$). In the exact algorithm, the speed of evolution can be manipulated such that the target is reached exactly at $2p=4$, i.e. after two iterations of $U(-\theta) U(\theta)$.}
    \label{fig:compare-approx-exact}
\end{figure*}

The Grover algorithm can be made deterministic with a variety of different approaches. One of the simplest approaches is via Long's algorithm \cite{Long2001}, which very slightly slows down the rate of amplification so that after an integer number of iterations the overlap with the marked state(s) reaches unity. We show that a similar approach is possible on CIINs. To our knowledge, this is the first example of deterministic quantum search not using the generalised rotation about the equal superposition $e^{-i \gamma \ket{s}\bra{s}}$ (or equivalently, quantum walk on the complete graph). Here, for convenience, we restrict to the case where $n \bmod 4 = 0$ so that $\nint{\frac{n}{4}} = \frac{n}{4}$. In this case, the walk times in the previous section simplify to $t_1 = \frac{\pi}{2}$ and $t_2 = \frac{\pi}{n}$. We then parametrise our previous iterate $U$, using an arbitrary phase angle $\theta$ to control the rotation speed,
\begin{equation}
    U(\theta) = U_w(\frac{\pi}{n}) U_f(\theta) U_w(\frac{\pi}{2}) U_f(\theta) \, .
\end{equation}
We will show that by applying $\left(U(-\theta) U(\theta)\right)^p$ to the initial state, with
\begin{equation}
    \theta = 2 \arcsin\left(\frac{n}{2 \sqrt{n-1}}\sin \left(\frac{1}{2 p}\arccos\frac{1}{\sqrt{n}}\right)\right)
    \label{eq:paramexact}
\end{equation}
where $p$ is the desired integer number of iterations, the entangled state is reached exactly. As a sidenote, in \cref{eq:paramexact} one cannot simply set $p=1$ to solve the search problem in a single iteration: in order for $\theta$ to be real-valued, the requirement on $p$ is that
\begin{equation}
    p \geq \frac{\arccos\left(\frac{1}{\sqrt{n}}\right)}{2 \arcsin\left(\frac{2 \sqrt{n-1}}{n}\right)} \approx \frac{\pi}{8}\sqrt{n} \, .
\end{equation}

To prove these dynamics, first substituting the definition of the above sequence of operators and simplifying, $U(-\theta)U(\theta)$ again rotates in a 2-dimensional subspace $\spn\{\ket{b_1^*}, \ket{b_4^*}\}$\, ,
\begin{equation}
    U(\theta) = 
\begin{pmatrix}
 \frac{e^{-i \theta }+n-1}{n} & \frac{e^{-i \theta } \left(-1+e^{i \theta }\right) \sqrt{n-1}}{n} \\
 -\frac{e^{-i \theta } \left(-1+e^{i \theta }\right) \sqrt{n-1}}{n} & -\frac{e^{-i \theta } \left(e^{i \theta }+n-1\right)}{n}
\end{pmatrix} \, .
\end{equation}
It is straightforward to verify the eigenphases of $U(-\theta) U(\theta)$ as
\begin{equation}
    \lambda_\pm = \pm 2 \arcsin\left( \frac{2 \sqrt{n-1}}{n} \sin\frac{\theta }{2}\right)
\end{equation}
with eigenstates
\begin{equation}
    \frac{1}{\sqrt{2}} \left( \ket{b_4^*} \pm e^{- i \gamma} \ket{b_1^*}\right)
\end{equation}
where $\gamma= \arctan\left(\frac{n-2}{n}\tan\frac{\theta }{2}\right)$.
This means that again, the system is rotated between the initial state and the fourth eigenstate of the adjacency matrix. However, in this case, the speed of rotation is controlled by the $\theta$ parameter. Using the eigensystem to find the state of the system after $p$ iterations of $U(-\theta) U(\theta)$ gives
\begin{equation}
    (U(-\theta) U(\theta))^p \ket{s} = \cos p \lambda_+ \ket{s} + i e^{- i \gamma } \sin p \lambda_+ \ket{b_4^*}
\end{equation}
and thus with $p = \frac{\arccos{\frac{1}{\sqrt{n}}}}{\lambda_+}$ the state
\begin{equation}
    \frac{1}{\sqrt{n}} \ket{s} + i e^{- i \gamma} \sqrt{\frac{n-1}{n}} \ket{b_4^*}
\end{equation}
is reached. Note that solving for $\theta$ gives \cref{eq:paramexact}. As before, to tune the phase difference and reach $\ket{+_\omega}$ exactly, a final quantum walk for time $t_3 = \frac{\pi}{2n} - \frac{\gamma}{n}$ is applied.

The final step, as before, is to map $\ket{+_\omega} \mapsto \ket{\omega}$. Here, we show that two phase queries are sufficient to perform the deterministic mapping from entangled to marked. It is easiest to work in the reverse direction, starting in the marked state (the conjugate transpose of a phase-walk evolution is simply the operators in reverse order with all parameters negated).
Consider the following evolution
\begin{equation}
    U_w(\frac{2 \pi j}{n}) U_f(\phi) U_w(\frac{2 \pi k}{n}) \ket{\omega}
\end{equation}
with $j, k \in \mathbb{Z}$. Calculating the resultant state gives (dropping the global phase),
\begin{align*}
    \cos\frac{\phi}{2}\left( \cos\frac{2(j+k)\pi}{n}\ket{\omega} - i \sin\frac{2(j+k)\pi}{n}\ket{\tilde{\omega}}\right) \\
    - \sin\frac{\phi}{2}\left( i \cos\frac{2(j-k)\pi}{n}\ket{\omega} + \sin\frac{2(j-k)\pi}{n}\ket{\tilde{\omega}}\right) \, .
\end{align*}
In order to have equal probability in each state, the parameters $(\phi, j, k)$ must then satisfy the requirement
\begin{equation}
    \cos ^2\frac{\phi }{2} \cos\frac{4 \pi  (j+k)}{n}+\sin ^2\frac{\phi }{2} \cos\frac{4 \pi  (j-k)}{n}=0
\end{equation}
Solving for $\phi$ in general,
\begin{equation}
    \phi = 2\arctan\sqrt{-\frac{\cos \left(\frac{4 \pi  (j+k)}{n}\right)}{\cos \left(\frac{4 \pi  (j-k)}{n}\right)}} \, .
\end{equation}
Thus for a given $n$, to obtain a real-valued phase rotation we need to find an integer pair $j$ and $k$ such that the expression inside the square root is non-negative. This is always possible: one choice that satisfies this requirement  for all $n \geq 8$ is $j = \floor{n/8}$ and $k=\ceil{n/8}$. This maps $\ket{\omega} \mapsto \frac{1}{\sqrt{2}}\left( \ket{\omega} + e^{i \gamma}\ket{\Tilde{\omega}}\right)$ with
\begin{equation}
    \gamma = \arccot\sqrt{\left(\frac{\sin \left(\frac{4 \pi  j}{n}\right)}{\cos \left(\frac{4 \pi  k}{n}\right)}\right)^2-1} \, ,
\end{equation}
and so a final phase shift $U_w(-\gamma)$ can be applied to eliminate the phase difference and obtain the Bell-like state $\ket{+_\omega}$ exactly. This gives a deterministic mapping between the marked and entangled states.

Hence, we obtain a deterministic algorithm to find the marked vertex on a CIIN for any value of $N$. Since this reduces to the prior algorithm when $\theta = \pi$, the algorithmic complexity is the same, requiring approximately $\frac{\pi}{2\sqrt{2}}\sqrt{N}$ oracle queries. A simulation of the approach is shown in \cref{fig:compare-approx-exact} for a small 24-vertex CIIN.

\subsection{A different path}
\label{sec:diffpath}

\begin{figure}
    \centering
    \includegraphics[width=0.9\linewidth]{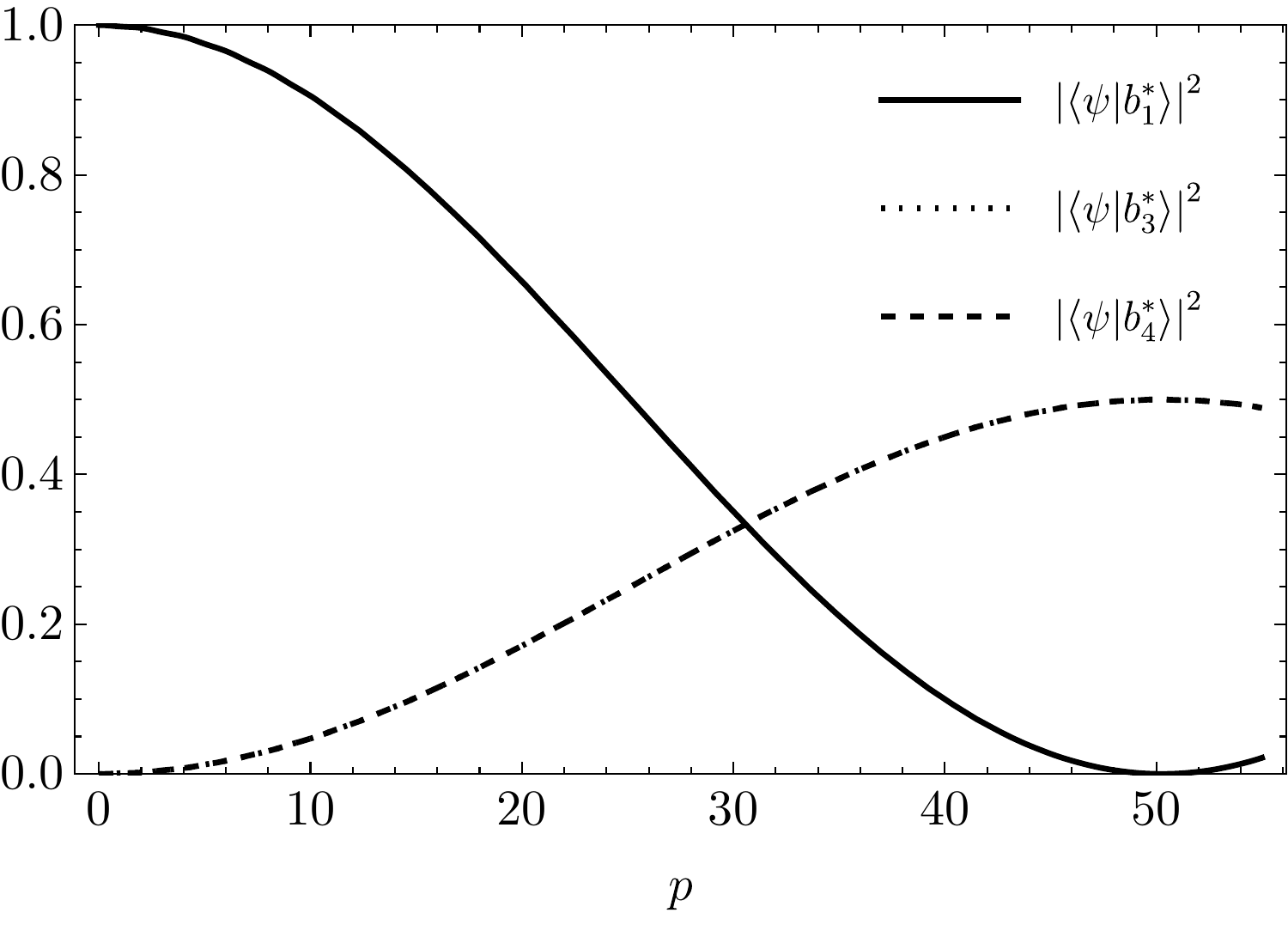}
    \caption{The state of a $2050$-vertex CIIN after $p$ applications of the alternate iterate $U_o$.}
    \label{fig:evolution-route2}
\end{figure}

We now demonstrate the flexibility of the framework by showing a different `path' through the search subspace that leads to the marked element, while still retaining the quadratic quantum speedup \textit{and} 100\% success probability. Here, assume that $n$ is odd. Then define the simple iterate
\begin{equation}
    U_o = U_w(\frac{\pi}{2}) U_f(\pi) \, .
\end{equation}
In the dual basis, $U_o^2$ takes the form
\begin{equation}
U_o^2 = \begin{pmatrix}
 \frac{n-2}{n} & 0 & \alpha  & \beta  \\
 0 & \frac{n-2}{n} & \beta  & \alpha  \\
  -i^n\alpha  & i^n \beta  & \frac{n-2}{n} & 0 \\
  i^n\beta  & -i^n\alpha & 0 & \frac{n-2}{n} \\
\end{pmatrix}
\end{equation}
where $\alpha = \left(-1+i^n\right) \frac{\sqrt{n-1}}{n}$ and $\beta = \left(1+i^n\right) \frac{\sqrt{n-1}}{n}$.
It immediately follows by observation that a rotation is induced in the subspace spanned by the initial state $\ket{b_1^*}$ and
\begin{equation}
    \ket{\xi} = \frac{(1 + i^n)}{2}\left(\ket{b_3^*} + i^n \ket{b_4^*}\right) \, .
\end{equation}
Expressing $U_o^2$ in the $\{ \ket{b_1^*}, \ket{\xi} \}$ basis gives
\begin{equation}
 U_o^2 = \begin{pmatrix}
\frac{n-2}{n} & -\frac{2 \sqrt{n-1}}{n} \\
 \frac{2 \sqrt{n-1}}{n} & \frac{n-2}{n} \\
\end{pmatrix} \, ,
\end{equation}
which is recognisable as the Grover iteration, with eigenvalues $\pm 2 \arcsin\frac{1}{\sqrt{n}}$ and eigenstates
\begin{equation}
    \frac{1}{\sqrt{2}}\left( \pm i \ket{b_1^*} + \ket{\xi} \right) \, .
\end{equation}
Hence, after $\frac{\pi}{2 \arcsin\frac{1}{\sqrt{n}}} \approx \frac{\pi}{4}\sqrt{n}$ iterations of $U_o^2$ the state $\ket{\xi}$ is reached. Now we observe that (again dropping the global phase)
\begin{equation}
    U_w(-\frac{\pi n}{4}) \ket{\xi} = \frac{1}{\sqrt{2}}\left( \ket{b_3^*} - \ket{b_4^*}\right) = \sqrt{\frac{n-1}{n}}\ket{b_1} - \frac{1}{\sqrt{n}} \ket{b_3}
\end{equation}
and thus this state has $\mathcal{O}(1)$ overlap with the marked state. The results of a numerical simulation on a $2050$-vertex CIIN (where $n=1025$) are shown in \cref{fig:evolution-route2}.

Note that this evolution, as before, can be made deterministic by slightly slowing the rotation rate such that the number of iterations required is integer-valued. We relegate the proof to  \cref{sec:proof-second-exact}, but one approach for determinism is
\begin{equation}
    U_o(\theta) = (U_w\left(\frac{\pi }{2}\right)U_f(-\theta))^2 (U_w\left(\frac{\pi }{2}\right)U_f(\theta))^2
    \label{eq:derand-2}
\end{equation}
where
\begin{equation}
    \ket{\xi} = U_o(\theta)^p \ket{b_1^*}
\end{equation}
with $\theta= 2 \arcsin\left(\frac{n}{2 \sqrt{n-1}}\sin\frac{\pi }{4 p}\right)$. Here, to make the angle $\theta$ real-valued, again $p \gtrapprox \frac{\pi}{8} \sqrt{n}$ is required.

To perform the final step and map completely to the marked state $\ket{\xi} \mapsto \ket{\omega}$, we follow a prescription very similar to the previous section, where $\ket{+_\omega}\mapsto \ket{\omega}$ was carried out. As before, for specific walk times $t$ the evolution can be constrained to the desired 2-dimensional subspace. In this case, when $n \bmod 2 = 1$ the choice $t=\pi$ induces rotation between $\ket{b_1}$ and $\ket{b_3}$:
\begin{align}
    U_w(\pi) (\alpha \ket{b_1} + \beta \ket{b_3}) = \frac{2\sqrt{n-1}}{n}(\beta \ket{b_1} + \alpha \ket{b_3}) \\ + \frac{n-2}{n}(-\alpha \ket{b_1} + \beta \ket{b_3}) \, .
\end{align}
Thus, a phase angle $\phi = 2 \arcsin\left(\frac{n^{3/2}}{4 (n-2) \sqrt{n-1}}\right)$ is chosen such that
\begin{align}
    U_w(\pi) U_f(\phi) U_w(\pi) \ket{b_1} = \sqrt{\frac{n-1}{n}}\ket{b_1} + \frac{e^{- i \gamma}}{\sqrt{n}}\ket{b_3} \, .
\end{align}
with $\gamma = \arctan\left(\frac{n^2 \cot \frac{\phi}{2}}{n^2-8 n+8}\right)$. After a final controlled phase shift $U_f(\gamma)$, we match up with the state reached in the previous section. Thus, overall we have shown that two efficient and unique rotations through the search space are possible, with both having 100\% theoretical success probability of measuring the marked element after $\frac{\pi}{2\sqrt{2}}\sqrt{N}$ iterations.

\section{Quantum circuit implementation}
\label{sec:circuit}

\begin{figure}
    \centering
    \includegraphics[width=0.9\linewidth]{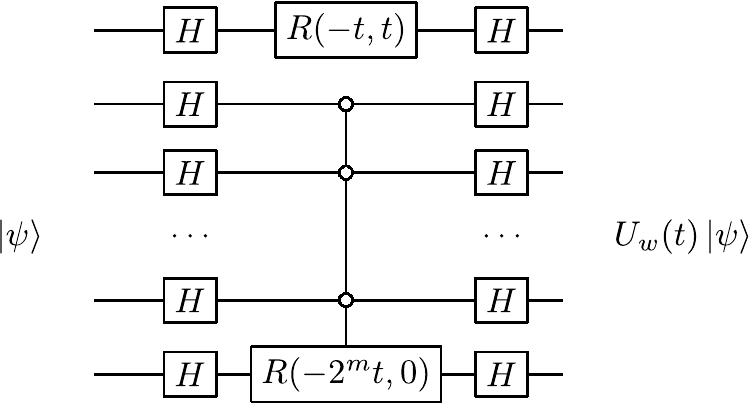}
    \caption{Quantum circuit for fast-forwarded gate-model simulation of quantum walk on a $2^{m+1}$-vertex CIIN.}
    \label{fig:quantum-circuit}
\end{figure}

In practice, an efficient spatial search algorithm also requires an efficient physical implementation of the associated quantum walks~\cite{QWbook2014}, for example, 
through an efficient quantum circuit~\cite{Nielsen2010, Douglas2009, Qiang2016efficient, Zhou2017}. Since the time-evolution operator of a CTQW on an unperturbed CIIN can be fast-forwarded, i.e. the quantum walk $U_w(t)$ can be simulated in constant-time with respect to $t$~\cite{Loke2017}, this makes this search approach amenable for implementation as a gate model quantum algorithm. The quantum circuit to implement a quantum walk on a $2^{m+1}$-vertex CIIN, which consists of two complete graphs of size $2^m$ with identity interconnections, is given in \cref{fig:quantum-circuit}. The rotation gate used in the circuit is defined as the generic two-phase rotation gate \cite{Loke2017},
\begin{equation}
    R(\theta, \phi) = \begin{pmatrix}
     e^{i \theta} & 0 \\
     0 & e^{i \phi}
    \end{pmatrix} \, .
\end{equation}
The first wire controls the identity interconnections, whilst the other $m$ wires represent the complete graph $\mathbb{K}_{2^m}$. Notably, in the quantum circuit diagram, the last $m$ wires of the circuit are exactly the generalised Grover diffusion operator, where a circuit for comparison can be found in \cite{Yoder2014}. As with ordinary Grover search, if one wishes to implement a quantum circuit for a $2n$-vertex CIIN where $n$ is \textit{not} a power of two, the arbitrary-modulus Quantum Fourier Transform $\mathcal{F}_n$ can be used to efficiently diagonalise $\mathbb{K}_n$ and replace the $\mathbb{K}_{2^m}$ component of the circuit \cite{Cleve2000, Marsh2020}. In reality, however, it is more convenient to round up the database size.

Hence, this work also provides motivation to develop fast-forwarded quantum circuits for continuous-time quantum walks on other graphs, in order to study the corresponding gate-model database search algorithm.

\section{Discussion and conclusion}
\label{sec:discussion}

In this work, we have illustrated the benefits of a novel spatial search algorithm for finding a marked vertex on a particular class of interdependent networks. The algorithm interleaves the two components of the spatial search Hamiltonian, performing alternating controlled phase shifts of the marked element followed by continuous-time quantum walks. We demonstrate that deterministic search can be achieved, and the number of oracle queries $\frac{\pi}{2\sqrt{2}}\sqrt{N}$ is comparable to Grover's algorithm for unstructured database search. We also observe that the walk and phase parameters do not converge to zero even as the number of iterations approaches infinity, indicating that this is not simply a Trotterised discretisation of standard spatial search.

The approach can be implemented as both a gate-model algorithm (c.f. Grover's algorithm) and an analog search algorithm (c.f. $\mathcal{CG}$ spatial search). Consequently, an interesting property of the framework is that the efficiency can be studied both in the query model and in terms of the total walk time. To compare to Grover's algorithm, the number of calls to $U_f(\theta)$ is summed. To obtain the total `evolution time' $T$ spent quantum walking, as per spatial search, the walk times is totalled. Quadratic quantum speedup is attained for CIINs in both cases, and time-efficiency would automatically be attained for any other $2\pi$-periodic graph that has $\mathcal{O}(\sqrt{N})$ query complexity. As an additional note, we also observe that there is the potential for some (necessarily dense) graphs to not have an efficient quantum circuit implementation for $U_w(t)$, and yet admit efficient gate-model search in terms of query complexity.

Although in this work the main focus is on a particular class of interdependent networks, we stress that this approach appears to work more generally on undirected graphs that admit efficient spatial search. Our work motivates the study of alternating phase-walk versions of spatial search on other graphs. The general pattern is to use the dual basis to compose phase-walk iterations that sequentially restrict to smaller subspaces. Although it is not immediately clear how to determine appropriate walk time and phase shift parameters given an arbitrary graph, there appears to be a strong connection to the spectrum of the adjacency matrix in terms of perfect state transfer and graph periodicity \cite{Godsil2011}. 

\acknowledgements

This research was supported by a Hackett Postgraduate Research Scholarship and an Australian Government Research Training Program Scholarship at The University of Western Australia. We thank Leonardo Novo for valuable insight and suggestions, and Lyle Noakes for his continuous support and discussions.



\bibliography{refs.bib}

\appendix

\section{Proof of reduced adjacency matrix}
\label{sec:reduced-adj}

Here, we prove the form of the adjacency matrix in the reduced walk basis. The reduction starts with
\begin{equation}
    A_\text{full} = \begin{pmatrix}
     \mathbb{K}_n & \mathbb{I} \\
     \mathbb{I} & \mathbb{K}_n
    \end{pmatrix} \,
\end{equation}
where $\mathbb{K}_n$ is the adjacency matrix of the complete graph, i.e. an all-ones matrix with zeroes on the diagonal. We primarily use the property that for $0 \leq x < n$
\begin{equation}
    A_\text{full} \ket{x} = \ket{n + x} + \sum\limits_{\substack{j=0\\j \neq x}}^{n-1} \ket{j}
\end{equation}
and for $n \leq x < N$
\begin{equation}
    A_\text{full} \ket{x} = \ket{x - n} + \sum\limits_{\substack{j=n\\j \neq x}}^{N-1} \ket{j} \, .
\end{equation}
It is sufficient to determine the action on $\ket{b_1}$ and $\ket{b_3}$, with the action on the other two basis states following by symmetry.
Hence,
\begin{align}
    A_\text{full} \ket{b_1} &= \sum\limits_{j=0}^{n-1} \ket{j} + \ket{n} = \sqrt{n-1} \ket{b_3} + \ket{b_2} \\
    A_\text{full} \ket{b_3} &= \frac{1}{\sqrt{n-1}} \sum\limits_{j=1}^{n-1} A_\text{full}\ket{j} \\
    &= \frac{1}{\sqrt{n-1}} \sum\limits_{j=1}^{n-1}\left(\sum\limits_{\substack{j=0\\j \neq x}}^{n-1} \ket{j} + \ket{n + x} \right) \\
    &= \frac{1}{\sqrt{n-1}} \sum\limits_{j=1}^{n-1}\left( \ket{0} - \ket{j} + \sum\limits_{\substack{j=1}}^{n-1} \ket{j} \right) + \ket{b_4} \\
    &= \sqrt{n-1} \ket{b_1} - \ket{b_3} + (n-1) \ket{b_3} + \ket{b_4} \\
    &= \sqrt{n-1} \ket{b_1} + (n-2) \ket{b_3} + \ket{b_4}
\end{align}
and thus the adjacency matrix takes the form as shown in \cref{eq:adjwalkbasis}.

\section{Proof of second approach to deterministic search}
\label{sec:proof-second-exact}

We use the iteration as per \cref{eq:derand-2}. First, define the following basis:
\begin{align}
    \ket{c_1} &= \ket{b_1} \, ,\\
    \ket{c_2} &= \frac{1}{2} \left(-1+i^n\right) \ket{b_3} + \frac{1}{2} \left(1+i^n\right) \ket{b_4} \, .
\end{align}
In this basis the reduced iterate takes the following form:
\begin{equation}
\begin{pmatrix}
 \frac{4 (n-1) \cos (\theta )+(n-2)^2}{n^2} & \frac{2 e^{-i \theta } \left(-1+e^{i \theta }\right) \sqrt{n-1} \left(e^{i \theta }+n-1\right)}{n^2} \\
 \frac{2 e^{-i \theta } \left(-1+e^{i \theta }\right) \sqrt{n-1} \left(1+e^{i \theta } (n-1)\right)}{n^2} & \frac{4 (n-1) \cos (\theta )+(n-2)^2}{n^2} \\
\end{pmatrix}
\end{equation}
which again can be verified to have eigenphases $\lambda_\pm = \pm 2 \arcsin\left( \frac{2\sqrt{n-1}}{n}\sin\frac{\theta}{2}\right)$ and corresponding eigenstates
\begin{align}
    \frac{1}{\sqrt{2}}\left( \pm e^{-i \arctan\left(\frac{(n-2)}{n}\tan\frac{\theta}{2}\right)}\ket{c_1} + \ket{c_2} \right) \, .
\end{align}
Thus, using this diagonalisation to compute the matrix power,
\begin{equation}
    U(\theta)^p \ket{c_1} = \cos\left( p \lambda_+\right) \ket{c_1} + i e^{i \arctan\left(\frac{(n-2)}{n}\tan\frac{\theta}{2}\right)} \sin \left( p \lambda_+\right) \ket{c_2} \,.
\end{equation}
Hence, with $p = \frac{\pi}{\lambda_+}$ the system is mapped to $\ket{c_2}$. Solving to find $\theta$ in terms of $p$ gives
\begin{equation}
    \theta =2 \arcsin\left(\frac{n}{2\sqrt{n-1}} \sin \left(\frac{\pi }{2 p}\right)\right)
\end{equation}
as required.
\vfill
\end{document}